# Correlation and recombination heating in an ultracold plasma. Analitic estimations.

Anatoli P. Gavrilyuk

*Institute of Computational Modeling of Siberian Branch of Russian Academy of Sciences*

*Krasnoyarsk 660036, Russia*

**Abstract.** In the article an ultracold electron-ion plasma created by photoionization of cooled atoms is investigated. We obtained analitical expressions for non-ideality parameters which establish due to correlation heating. In the work the nearest neighbour and the Wigner-Seitz cell approximations were used, the recombination heating of electrons was taken into account. We have got a good agreement with the experiment results. The possibility of creation of strongly non-ideal electron subsystem has been shown and conditions of this process have been determined.

The experements [1,2] resulted in growing interest to an ultracold plasma (UP) with temperatures of particle $T_{e,i} \leq 100\ K$. In these works a quasineutral ultracold electron-ion plasma was first obtained by means of a nearthreshold photoionization of cooled atoms. It was expected that by this way strongly non-ideal plasma could be produced with $\Gamma_e$, $\Gamma_i \gg 1$, where $\Gamma_e$ and $\Gamma_i$ are the parameters of non-ideality of electron and ion subsystems respectively. But in the posterior works [3-6] based on using of molecular dynamics and Monte Carlo techniques it was shown that the particles formed due to photoionization were heating up quickly. This so-called correlation heating (or disorder-induced heating) is caused by establishing of spatial distribution of particles corresponding to the minimum potential energy. The last means that the degree of correlation of spatial distribution grows fast in initially non-correlated particle ensemble. The second impotant cause is heating of electrons at a three-body recombination. The causes lead to significant decrease of non-ideality of both electron and ion subsistems even before the plasma cloud begins to broaden out.

The aim of the work is to get rather simple and adequate expressions describing the above mentioned processes and the state of plasma at times $t \leq \omega_e^{-1}$, $\omega_i^{-1}$ ($\omega_e$, $\omega_i$ – the electron and ion plasma frequencies respectively). It would make much easy to analyze the possibility and conditions of creation strongly non-ideal UP.

### Correlation and recombination heating of electrons.

When UP is created by photoionization there are two main processes (besides the photoionization) determining the temperature of electrons: correlation heating and three-body recombination.

Let's consider first the correlation heating of the electron subsistem and estimate the temperature and the parameter of non-ideality $\Gamma_e$, which establish in UP produced by fast photoionization of cold atoms. The non-ideality parameter is given by the expression:

$$\Gamma = \frac{z^2 e^2}{a}/kT, \qquad \frac{4}{3}\pi a^3 n = 1, \qquad (1)$$

where e – the elementary charge, $z_e$ – the particle charge (further $z_e = 1$), a – the Wigner-Seitz radius, k – the Boltzmann constant, T – the temperature of particles, n – the concentration of particles.

So cooled atom gas with negligibly small kinetic energy of the particles is exposed to photoionizing radiation with photon energy $\hbar\omega = I_0 + \Delta$, $I_0$ – the ionization potential of atom, $\Delta$– the initial kinetic enrgy of formed free electrons and $I_0 \gg \Delta$. Thus when photoionizing the atom, the energy $\hbar\omega$ is added to its initial energy ($-I_0$) and the energy of plasma becomes equal to

$$E_p = -NI_0 + NI_0 + N\Delta = N\Delta, \qquad (2)$$

where N – the full number of ions produced. From the other side, neglecting the motion of ions the Hamiltonian of this plasma (classical) can be written as

$$H_p = \frac{1}{2}\sum_{k=j}^{N}\frac{e^2}{|r_k^e - r_j^e|} + \frac{1}{2}\sum_{k \neq j}^{N}\frac{e^2}{|r_k^i - r_j^i|} - \sum_{k \neq j}^{N}\frac{e^2}{|r_k^i - r_j^e|} + \sum_{k}^{N}\left(-\frac{e^2}{|r_k^i - r_k^e|} + \varepsilon_{ek}\right), \qquad (3)$$

where $r_k^e$, $r_k^i$ – the radius-vectors of the k-th electron and ion respectively, $\varepsilon_{ek}$ – the kinetic energy of the k–th electron. The first three sums describe the Coloumb interaction energy. In the last sum the first term is the energy of Coloumb interaction of the ion with the nearest electron., For the nearest neighbour approximation (ignoring the interaction between the charged particle and all other particles except the nearest one with the opposite charge) the total contribution of the first three terms is zero.

Thus the last sum gives the main contribution. Averaging over all particles, we have for the plasma energy:

$$E_p = N\left\langle -\frac{e^2}{r^i - r^e}\right\rangle + N\langle\varepsilon_e\rangle = N\cdot\Delta, \qquad (4)$$

where $\left\langle -e^2/|r^i - r^e|\right\rangle = \langle U_{ei}\rangle$, $\langle\varepsilon_e\rangle$ - the mean values of the electron-ion interaction energy and the electron kinetic energy, respectively. The first one we estimate to the nearest neighbour approximation (as in [7-9]). The density of probability of finding the nearest electron on the distance $R = |r^i - r^e|$ from the ion is

$$F(R) = \frac{3}{a^3} R^2 \exp\left(-\frac{R^3}{a^3}\right), \qquad (5)$$

and the energy of their interaction is $U_{ei} = -e^2/R$. Using (5) we get an estimation:

$$\langle U_{ei} \rangle \approx -\int_0^\infty \frac{e^2}{R} \cdot \frac{3}{a^3} R^2 \exp\left(-\frac{R^3}{a^3}\right) dR = -\frac{e^2}{a} \Gamma am\left(\frac{2}{3}\right) \approx -1.35 \frac{e^2}{a}. \qquad (6)$$

Here $\Gamma am$ – the gamma-function. From (4) and (6) we get

$$\langle \varepsilon_e \rangle = \Delta - \langle U_{ei} \rangle = \Delta + 1.35 \frac{e^2}{a}. \qquad (7)$$

Then the non-ideality parameter of the electron subsystem $\Gamma_e$ is equal to

$$\Gamma_e = \frac{e^2/a}{(2/3)[\Delta - \langle U_{ei} \rangle]} = \frac{e^2/a}{(2/3)[\Delta + 1.35 \cdot (e^2/a)]}. \qquad (8)$$

When $\Delta \geq 0$, the maximum value of the non-ideality parameter is

$$\Gamma_e = \frac{\Gamma_0}{1 + 0.9\Gamma_0} \leq 1.11, \qquad \Gamma_0 = \frac{e^2}{a} \bigg/ \frac{2}{3}\Delta, \qquad (9)$$

where $\Gamma_0$ – the non-ideality parameter, formally calculated for the initial energy of photoelectrons $\Delta$. Though the value of $\Gamma_0$ can be as big as is wished, one can see from (9) that when relaxing the electrom subsystem, the real non-ideality parameter establishes at the level $\Gamma_e \sim 1$, that is the electrons can't be strongly non-ideal at $\Delta > 0$. The time of relaxation $\tau_e \sim \omega_e^{-1}$, $\omega_e$ – the electron plasma frequency. This result is in a good agreement with the data [5] obtained on basis of molecular dynamics method.

Moreover it can be seen from (8) that it is possible to reach the range $\Gamma_e \gg 1$ only if

$$0 > \Delta > \langle U_{ei} \rangle. \qquad (10)$$

Under this condition the average kinetic energy of electrons $\langle \varepsilon_e \rangle > 0$ (Fig.1), they are "delocalized" [9]. These can be outermost electrons of Ridberg atoms with crossing areas of localization.

At $\Delta < \langle U_{ei} \rangle$ we have $\langle \varepsilon_e \rangle < 0$ from (7) that contradicts the phisical sense. Indeed under this condition free ("delocalized") electrons are absent. So the creation of strongly non-ideal electron subsystem is possible only if the condition (10) is satisfied.

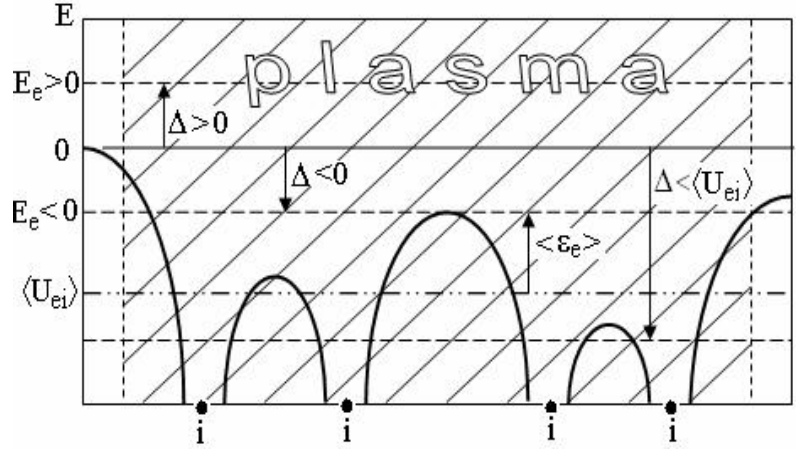

Fig. 1. The diagram for $E_e$ energies in plasma: when $\Delta < \langle U_{ei} \rangle$ there are no free electrons, points i – the ions.

To estimate an influence of heating due to the three-body recombination on the electron temperature we will use the description of recombination suggested in [10]. According to the article the recombination is treated as a diffusion of electrons on atom energy states and is described by the Fokker-Planck equation. To a stationary ( on the energy states) recombination flow approximation it can be written

$$\frac{dn}{dt} = -j = -\frac{1,2 \cdot 10^{-7} n^3}{(\varepsilon_e / k)^{9/2}}, \qquad (11)$$

where $\varepsilon_e$ – the average kinetic energy of free electrons. We assume that its initial value is defined by (7). At this approach the recombination represents as a stepped sequence of electron transitions between bound states due to collisions with free electrons. When the bound electrons shift down the energy axis their energy is transmitted to free electrons, the recombination heating of the last occurs. Besides the stepped process resulting in the heating of free electrons there are down transitions due to spontaneous decay. For high excited (Ridberg) states the rate of such process is small in comparison with the stepped one but when lowering of a state the stepped process rate decreases and the rate of the spontaneous one increases. The boundary energy $E_R$ can be defined [11]

$$E_R / k = \frac{15 \cdot n^{1/4}}{(\varepsilon_e / k)^{1/8}} (K), \qquad (12)$$

below which the spontaneous decays dominate. So it can be considered that when the recombination of one electron occurs the free electrons are given the energy $W_r = E_R + \varepsilon_e$. Then to a stationary recombination flow approximation the change of kinetic electron energy is described by the equation:

$$\frac{d\varepsilon_e}{dt} = \frac{E_R + \varepsilon_e}{n} j. \tag{13}$$

As at $n \sim 10^6 \div 10^9$ cm$^{-3}$ and $\varepsilon_e/k < 100$ K, $E_R/k \sim 300 \div 2500$ K so we neglect $\varepsilon_e$ in (13). We neglect also the change n during the time $t_r$. After substitution for the expressions for j and $E_R$ in (13) we get:

$$\frac{d(\varepsilon_e/k)}{dt} = \frac{1,8 \cdot 10^{-6} n^{2.25}}{(\varepsilon_e/k)^{4.625}}. \tag{14}$$

Integrating (14) from $t = t_0$ to $t = t_r$ we have:

$$\varepsilon_e/k \approx \left[10^{-5} n^{2.25}(t_r - t_0) + (\varepsilon_0/k)^{5.625}\right]^{0.178}, \tag{15}$$

where $\varepsilon_0$ – the initial kinetic electron energy which can be equal to $\langle \varepsilon_e \rangle$ in (7). It is easy to see that when $\varepsilon_0/k < \left[10^{-5} n^{2.25}(t_r - t_0)\right]^{0.178}$ the kinetic energy $\varepsilon_e$ depends weakly on its initial value $\varepsilon_0$. Experimental results from [2] can be used to make sure in adequacy of (15). In the figure 2 taken from this work the experimental data for plasma clouds expanding velocities $v_0$ depending on $\varepsilon_0$ for different initial concentrations of Xe plasma are presented. On the figure there are also theoretical dependencies $v_0(\varepsilon_0)$ obtained from (15) and the following formular in [2]:

$$v_0 = \sqrt{\varepsilon_e(\varepsilon_0)/\alpha M}.$$

Here α=1.7 – a trimming parameter, M – the ion mass.

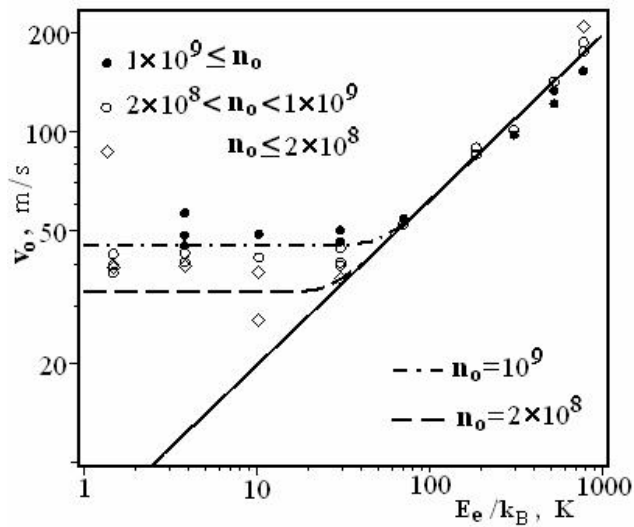

Fig.2 The velocity of plasma expansion $v_0$ for different initial concentrations (in cm$^{-3}$). Dashed lines – the calculations on basis of (15).

The calculation of $\varepsilon_e$ was done for $t_0 = 0$. The value $t_r$ was chosen the expansion of the cloud to be insignificant. As it follows from [2] (see Fig.2) this time is $t_r \approx 3$ мкс. It is seen from the Fig.2 that there is a good agreement between the theoretical and experimental data.

Now using (15) we can estimate the energy of electrons and their non-ideality parameter which establish for time not eexceeding the relaxation time of the ion subsystem $\tau_i \sim \omega_i^{-1}$ when shift of ions is insignificant. Disregarding $t_0 \sim \omega_e^{-1}$ (in comparison with $\tau_i$) and $\varepsilon_0 / k$ (at $\Delta=0$) in (15) we get:

$$(\varepsilon_e / k)_{min} \approx 0.13 \cdot n^{0.4} \tau_i^{0.178} = 0.036 \cdot n^{0.31} A^{0.09}, \quad (16)$$

where A is the atom mass of the ion. Then the non-ideality parameter takes on a value

$$(\Gamma_e)_{max} \approx 0.11 \frac{n^{0.02}}{A^{0.09}}. \quad (17)$$

Thus the maximum of $\Gamma_e$ which establishes in the electron subsystem before ions start moving is weakly depends on the plasma concentration and the ion mass and is approximately equal to $\Gamma_e \sim 0.1$. Note that $\Gamma_e$ decreases when $\Delta$ (or $\varepsilon_0$) grows.

It is easy to make sure that the quantity ($\Delta n$) of recombined at this process electrons is a small part of initially formed ones. Since the energy emitted under recombination is equal to the energy recieved by the free electrons we have:

$$E_R \cdot \Delta n = (\varepsilon_e - \varepsilon_0) n. \quad (18)$$

The recombination speed is maximum when the kinetic energy is minimum described by (16). Then from (18) and (16) we get

$$\frac{\Delta n}{n} = \frac{\varepsilon_e}{E_R} \approx 1.6 \cdot 10^{-3} (nA)^{0.1}. \quad (19)$$

Even if $n = 10^{10}$ cm$^{-3}$ and $A = 100$ it leads to only $\Delta n / n \approx 0.025$. That is the relative part of recombined electrons (ions) for the time $t \leq \omega_i^{-1}$ is really very small. This fact validates the disregarding of change of concentration used above under the conditions considered.

**The correlation heating of ions.**

As it follows from the preceding results, after photoionization the electron subsystem rapidly becomes weakly non-ideal. At the same time the initial spatial distribution of formed ions is homogeneous and random. That is for times $t < \tau_i$ the ion-ion correlation is absent. Taking into account the weak non-ideality of electrons we can neglect both electron-electron and electron-ion correlations. In this case the ensemble average energy of electrostatic interaction per one ion

$\langle U_{i0}^{es}\rangle$ equals zero. Later the relaxation of spatial ion distribution to more odered one occurs resulting in growing of kinetic energy of ions (the correlation heating). To estimate the magnitude of this heating we use a one component plasma model (OCP). The role of background which nuetralizes ions is played by the weakly non-ideal electron subsystem. Note that the correlation heating of ions in OCP for initially disordered spatial distribution was first considered in [12].

To determine the mean kinetic energy of ions they have after the correlation heating we use an approximation of "ion spheres" [13] or Wigner-Seitz cell (Fig.3). The approximation is valid if the non-ideality parameter of ions establishes in the range $\Gamma_i > 1$ (this is true as we will see below). The charge of uniform background of the cell is equal to the ion charge with opposite sign.

The electrostatic energy of interaction between ion and background of the cell is [13]:

$$U_i = -\frac{3}{2}\cdot\frac{e^2}{a}+\frac{e^2}{2a}\cdot\left(\frac{r}{a}\right)^2, \qquad (20)$$

where the first term is the energy of the ion when it is in the center of the cell and the second one takes account of the shift of the ion on the distance **r** from the cell center.

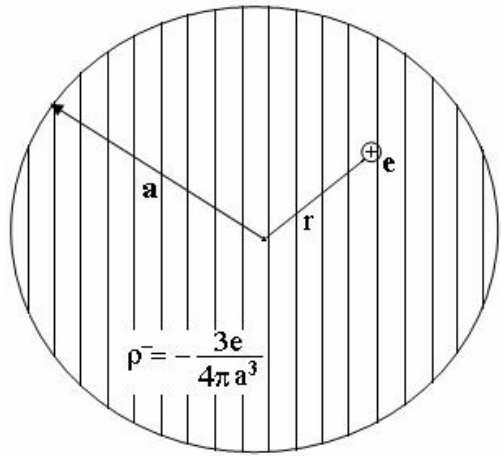

Fig.3. The model of "ion sphere". $\rho^-$ - the charge density of the uniform background, **r** - the distance between ion and the cell center.

The respective energy of the cell background is:

$$U_e = \frac{3}{5}\cdot\frac{e^2}{a}. \qquad (21)$$

Then the full energy of electrostatic interaction into the cell containing an ion equals to:

$$U^{es} = U_i + U_e = -0.9\frac{e^2}{a}+\frac{e^2}{2a}\cdot\left(\frac{r}{a}\right)^2. \qquad (22)$$

Note that the second term in the right side of (20) describes the potential energy of harmonic oscillator. Its averige on ensemble meaning is equal to the averige kinetic energy of ions:

$$\left\langle \frac{e^2}{2a} \cdot \frac{r^2}{a^2} \right\rangle = \left\langle \varepsilon_i \right\rangle = \frac{3}{2} kT_i . \qquad (23)$$

In UP [1,2] the initial kinetic energy of ions is negligibly small ($\varepsilon_{i0}/k \sim 1\,\mu K$) that is why we assume it is zero. The initial electrostatic energy is also zero. The system is closed, so the full energy is constant. Then we can write down for the mean full energy:

$$\left\langle E \right\rangle = \left\langle \varepsilon_i \right\rangle + \left\langle U^{es} \right\rangle = \left\langle \varepsilon_{i0} \right\rangle + \left\langle U^{es}_{i0} \right\rangle = 0 . \qquad (24)$$

Having regard to (22) and (23) the expression (24) takes on a form:

$$\left\langle E \right\rangle = \left\langle \overline{\varepsilon_i} \right\rangle - 0.9 \frac{e^2}{a} + \left\langle \overline{\varepsilon_i} \right\rangle = 3kT_i - 0.9 \frac{e^2}{a} = 0. \qquad (25)$$

From here we get the ion temperature and the non-ideality parameter those the correlation heating has led to

$$kT_i = 0.3 \frac{e^2}{a}, \qquad \Gamma_i = \frac{e^2/a}{kT_i} \approx 3.3 . \qquad (26)$$

Below in the table the theoretical (defined by (26)) and experimental temperatures of ions after the correlation heating are presented. For the theoretical temperature of electrons the lower limit obtained from (16) has been given. It is seen that the electron temperature in experiments exceeded the theoretical minimum. That is initially the electrons were weakly non-ideal and the role of three-body recombination for times $\sim \tau_i$ is negligible. The experimental temperatures of ions presented in the table relate mostly to the central part of the plasma cloud. In this area plasma is more uniform and its density is close to the peak value given in the table.

**The theoretical and experimental temperatures of ions Sr$^+$ when the correlation heating in UP has occured.**

| № | n, cm$^{-3}$ | T$_e$, K (theor.) | T$_i$, K (theor.) | T$_e$, K (exp.) | T$_i$, K (exp.) | source |
|---|---|---|---|---|---|---|
| 1 | $5 \cdot 10^9$ | ≥36.6 | 1.4 | 46 | 1.3 – 1.4 | [14] |
| 2 | $4 \cdot 10^9$ | ≥34.7 | 1.3 | 38±6 | 0.9 –1.2 | [15] |
| 3 | $1.5 \cdot 10^9$ | ≥25.2 | 0.94 | 37 | 0.9 - 1.1 | [16] |
| 4 | $5.7 \cdot 10^9$ | ≥38.1 | 1.46 | 56 | 1.8 -2.0 | [16] |

As we see from the table there is a good agreement between the theoretical and experimental values of ion temperature got as a result of the correlation heating.

**Conclusion.**

The estimations performed have allowed to obtain simple and adequate expressions for temperatures (non-ideality parameters) of electrons and ions those establish due to the correlation heating of the particles. These expressions are universal, they do not dependent on sort of ions, and to find the mentioned parameters we do not need to use any table values or to perform additional calculations. The discovered possibility to get a strongly non-ideal electron subsystem when exciting atoms to some specified Ridberg states is of particular interest. It can be expected that using of odered («cristall») initial atom distribution will be found most advantageous. It has been shown that the recombination heating is the major factor determining the electron temperature for times $\leq \tau_i$. The approach used [9] descriibes the process quite well. The recombination heating leads to the fact that motion of ions in UP is always going on against the background of weakly non-ideal electrons (namely this allows to use the OCP model). That is why using of UP models [6], where the strong shielding of ions by electrons ($a/\lambda_e > 1$, $\lambda_e$ - the electron Deby radius) is considered, is probably not valid. The expression derived for $\Gamma_i$ does not dependent on the temperature of electrons. However such dependence was obtain in the work [15]. It can be partly explained by shielding (even weak) of ions by electrons. It can also be reffered to the inhomogeneity of the plasma cloud which must lead to inhomogeneous recombination and correlation heating of the particles giving rise to additional flows of energy and particles.